\documentclass[reprint,twocolumn,aps,superscriptaddress,preprintnumbers,prd]{revtex4-2}

\usepackage[margin=1in]{geometry}
\usepackage[T1]{fontenc}
\usepackage[utf8]{inputenc}
\usepackage{lmodern}
\usepackage{amsmath,amssymb,bm}
\usepackage{graphicx}
\usepackage{booktabs,array}
\usepackage{xcolor}
\usepackage[colorlinks=true,linkcolor=blue,citecolor=blue,urlcolor=blue]{hyperref}
\usepackage{microtype}

\graphicspath{{figures/}}

\begin{document}
\title{Digital Quantum Simulation of Nonequilibrium Dynamics in the Schwinger Model under a Strong External Electric Field}
\author{Haobin Chen}
\affiliation{State Key Laboratory of Nuclear Physics and Technology, Institute of Quantum Matter, South China Normal University, Guangzhou 510006, China}
\affiliation{Key Laboratory of Atomic and Subatomic Structure and Quantum Control (MOE), Guangdong-Hong Kong Joint Laboratory of Quantum Matter, Guangzhou 510006, China}
\affiliation{Guangdong Basic Research Center of Excellence for Structure and Fundamental Interactions of Matter, Guangdong Provincial Key Laboratory of Nuclear Science, Guangzhou 510006, China}
\author{Lin Cheng} 
\affiliation{State Key Laboratory of Nuclear Physics and Technology, Institute of Quantum Matter, South China Normal University, Guangzhou 510006, China}
\affiliation{Key Laboratory of Atomic and Subatomic Structure and Quantum Control (MOE), Guangdong-Hong Kong Joint Laboratory of Quantum Matter, Guangzhou 510006, China}
\affiliation{Guangdong Basic Research Center of Excellence for Structure and Fundamental Interactions of Matter, Guangdong Provincial Key Laboratory of Nuclear Science, Guangzhou 510006, China}
\author{Xingyu Guo}
\email{guoxy@m.scnu.edu.cn}
\affiliation{State Key Laboratory of Nuclear Physics and Technology, Institute of Quantum Matter, South China Normal University, Guangzhou 510006, China}
\affiliation{Guangdong Basic Research Center of Excellence for Structure and Fundamental Interactions of Matter, Guangdong Provincial Key Laboratory of Nuclear Science, Guangzhou 510006, China}
\collaboration{QuNu Collaboration}

\begin{abstract}
We use the (1+1)-dimensional Schwinger model to investigate the nonequilibrium dynamics of a finite lattice system under a constant external electric field. The lattice Hamiltonian is constructed under open boundary conditions. The vacuum state is prepared using the variational quantum eigensolver (VQE). Scans over the external field strength show the flip of the vacuum state at several field strengths. The critical field strengths agree with theoretical predictions. We further investigate the real-time evolution of the zero-field vacuum under an external electric field using a second-order Trotter-Suzuki decomposition. By comparison with exact diagonalization (ED), we verify that the quantum-simulation protocol reproduces the main features of field-induced boundary charge separation, decay of the vacuum-state fidelity, and quasiperiodic energy redistribution between the electric-field energy term and the fermionic sector. Our results indicate that combining VQE-based state preparation with digital real-time evolution provides a useful approach for studying nonequilibrium dynamics in strong-field lattice gauge theories.
\end{abstract}

\maketitle


\section{Introduction}

Vacuum instability in strong-field quantum electrodynamics is an
important problem in nonperturbative quantum field theory. In a
sufficiently strong electric field, the quantum vacuum can decay through
tunneling and produce particle-antiparticle pairs, a process known as
the Schwinger effect \cite{ref1,ref2,ref3,ref4,ref5}. This process exhibits a characteristic nonperturbative exponential dependence and is intrinsically a real-time nonequilibrium phenomenon \cite{ref3,ref6,ref7,ref8}. It is therefore difficult to describe reliably using traditional perturbation theory. Numerical methods capable of directly tracking the real-time evolution of quantum states are essential for understanding vacuum structure, pair production, and energy transfer mechanisms in strong-field QED \cite{ref9,ref10,ref11,ref12}.

Lattice gauge theory provides a first-principles framework for studying
nonperturbative quantum field theory \cite{ref13,ref14,ref15,ref16,ref17}. In Euclidean path-integral Monte Carlo simulations, efficient importance sampling is possible when the weight can be interpreted as a non-negative
probability distribution. However, when the weight becomes complex,
non-positive, or highly oscillatory because of a chemical potential, a
topological term, a fermion determinant, or a real-time phase factor,
severe cancellations occur among different field configurations, and the statistical error grows rapidly with the system volume or the evolution
time. This difficulty is usually referred to as the sign problem \cite{ref18,ref19,ref20}. The Hamiltonian formalism is not by itself a source of quantum speedup, but it does not rely on Euclidean probability-weight sampling and can therefore bypass this class of sign problems associated with non-positive sampling weights. The corresponding difficulty is then shifted to the direct treatment of an exponentially large Hilbert space. Quantum computation and quantum simulation provide natural platforms for
real-time lattice field theory, since quantum many-body states and
unitary time evolution can be directly represented with qubits \cite{ref21,ref22,ref23,ref24}.

The Schwinger model, namely quantum electrodynamics in 1+1 dimensions, is a standard low-dimensional model for studying strong-field
nonequilibrium dynamics \cite{ref25,ref26}. Although it is much simpler than
QED in 3+1 dimensions, it retains several essential features, including
local U(1) gauge invariance, charge screening, confinement, chiral
condensation, and vacuum instability \cite{ref27,ref28,ref29,ref30,ref31,ref32,ref33,ref34,ref35,ref9,ref36,ref10}. In addition, gauge fields in one spatial dimension have no independent degrees of freedom. Under open boundary conditions, Gauss's law allows the electric field on each link to be expressed as a function of the matter charge distribution and the boundary electric field, thereby reducing the model to a purely fermionic Hamiltonian or, after a Jordan-Wigner
transformation, a qubit Hamiltonian \cite{ref29,ref30,ref37,ref38,ref39}. This feature makes the Schwinger model a useful benchmark platform for digital
quantum simulation of lattice theories \cite{ref40,ref41,ref42,ref43,ref44,ref45,ref46,ref47,ref48}.

In recent years, the Schwinger model has become an important test bed
for quantum simulations in high-energy physics. The real-time dynamics,
pair production, thermalization, phase structure, and finite-temperature properties of the Schwinger model and related lattice models
have been studied using platforms such as trapped ions,
superconducting qubits, tensor networks and cold atoms \cite{ref43,ref44,ref49,ref50,ref51,ref52,ref53,ref54,ref55,ref56,ref57,ref58,ref11,ref12,ref59,ref60,ref61}. Related analog and digital quantum-simulation studies have also investigated gauge invariance, thermalization, topological $\theta$ angles, confinement dynamics, scattering processes, and higher-dimensional gauge-theory dynamics \cite{ref62,ref63,ref64,ref65,ref66,ref67,ref68,ref69,ref70,ref71,ref72}. In particular, quenches in which an external electric field is suddenly switched on have been used to
investigate strong-field-induced particle-antiparticle production and
vacuum response \cite{ref9,ref11,ref12}. In related studies, the zero-field ground state is often prepared using VQE, evolved under the Hamiltonian
with an external field, and compared with results from exact diagonalization or tensor-network calculations \cite{ref9,ref12,ref73,ref74,ref75,ref76}.

In this work, we study the lattice Schwinger model with a constant
external electric field and focus on the finite-size nonequilibrium
dynamics after an electric-field quench. We first construct the lattice
Hamiltonian under open boundary conditions and map it to a
Pauli-operator form using the Jordan-Wigner transformation. We then use
VQE to prepare the zero-field vacuum state, which is taken as the
initial state when the external field is switched on. The real-time evolution is implemented by a second-order Trotter-Suzuki decomposition.
In contrast to approaches based on the McLachlan variational principle
and fixed-depth variational real-time evolution, the present work uses
explicit digital Trotter evolution and examines the total charge, total
energy, spatial-point charge distribution, vacuum-state fidelity, and
electric-field energy \cite{ref11,ref12,ref77,ref78,ref79,ref80,ref81}. By comparing these quantities with ED results, we assess the reliability of the protocol for
describing strong-field dynamics in finite lattice systems. The numerical implementation is based on the Qiskit quantum computing framework \cite{ref82,ref83}. We also perform scans over the external field and
analyze the chiral condensate, the low-energy levels, and the finite-size critical external field.

The paper is organized as follows. Section II introduces the lattice
Schwinger model with an external electric field, the Gauss-law reduction, and the qubit encoding. Section III describes the VQE state
preparation and the second-order Trotter-Suzuki evolution. Section IV
presents numerical results. Section V summarizes the work and discusses possible extensions, including larger system sizes, improved ansatzes, and
implementation on noisy quantum hardware. Appendix A gives dynamical
checks for the $N = 12$ lattice.

\section{Lattice Schwinger Model with an External Electric Field}

\subsection{Continuum Model and Lattice Discretization}

In natural units, the continuum Lagrangian density of the Schwinger
model is \cite{ref25,ref26,ref28}
\begin{equation}
    \mathcal{L} = \bar{\psi}\left( i\gamma^{\mu}D_{\mu} - m \right)\psi - \frac{1}{4}F_{\mu\nu}F^{\mu\nu},
\end{equation}

where $D_{\mu} = \partial_{\mu} + igA_{\mu}$ is the covariant
derivative, $\psi$ is the fermion field, $m$ is the bare fermion
mass, $g$ is the gauge coupling, and $A_{\mu}$ is the $U(1)$ gauge
field. {The field-strength tensor is defined as
$F_{\mu\nu} = \partial_{\mu}A_{\nu} - \partial_{\nu}A_{\mu}$.
In $1+1$ dimensions, the electric field is the only independent
component of the field-strength tensor, and we adopt the convention
$E \equiv F_{01} = \partial_{0}A_{1} - \partial_{1}A_{0}$.
In the temporal gauge $A_{0} = 0$, this reduces to
$E = \partial_{0}A_{1}$,} and the Hamiltonian can be written as
\begin{equation}
  H = \int dx\left[ - i\bar{\psi}\gamma^{1}D_{1}\psi + m\bar{\psi}\psi + \frac{1}{2}E^{2} \right].  
\end{equation}

To simulate this theory with a finite number of qubits, the continuum
field theory is discretized on a finite lattice using the
Kogut-Susskind staggered-fermion formulation \cite{ref14,ref29,ref30}.
Writing the continuum Dirac field as
$\psi(x)=(\psi_1(x),\psi_2(x))^T$, its two spinor components are
assigned to neighboring staggered lattice sites. The mapping can be
written as
\begin{equation}
\begin{pmatrix}
\psi_1(x_i)\\
\psi_2(x_i)
\end{pmatrix}
\longleftrightarrow
\frac{1}{\sqrt{a}}
\begin{pmatrix}
\phi_{2i-2}\\
\phi_{2i-1}
\end{pmatrix}.
\end{equation}
Here, $i=1,\ldots,N/2$ labels the physical spatial points, with each
point represented by two neighboring staggered lattice sites.
The fermionic degrees of freedom therefore reside on lattice sites,
whereas the gauge degrees of freedom reside on the links between
neighboring sites.

Consider a system with $N$ lattice sites and lattice spacing $a$,
with the sites labeled by {$n = 0,1,\ldots,N-1$}. The discretized lattice
Hamiltonian is

\begin{eqnarray}
    H &=& -\frac{i}{2a}\sum_{n = 0}^{N - 2}
    \left( \phi_{n}^{\dagger}U_{n}\phi_{n + 1}
    - H.c. \right) 
    \nonumber\\
    &&+ m\sum_{n = 0}^{N - 1}( - 1)^{n}
    \phi_{n}^{\dagger}\phi_{n}
    + \frac{g^{2}a}{2}\sum_{n = 0}^{N - 2}L_{n}^{2}.
\end{eqnarray}

Here, $\phi_{n}$ is the staggered-fermion operator on the $n$-th
lattice site, {$U_n \equiv U_{n,n+1}$ is the gauge-link variable on
the link connecting sites $n$ and $n+1$},
and $L_{n}$ is the corresponding lattice electric field. The first
term describes hopping between neighboring lattice sites, the second
term is the staggered mass term, and the third term is the
electric-field energy. Unless otherwise specified, all numerical
calculations in this work are performed in lattice units with
$a = m = g = 1$.

\subsection{Gauss's Law, Open Boundary Condition, and the External Electric Field}

In $1 + 1$ dimensions, the gauge field has no independent propagating
degrees of freedom, and the electric field on each link is fully
determined by Gauss's law and the boundary conditions. The lattice Gauss
law is \cite{ref14,ref29,ref30}

\begin{equation}
L_{n} - L_{n - 1} = \phi_{n}^{\dagger}\phi_{n} - \frac{1 - ( - 1)^{n}}{2}.
\end{equation}

{In this work, we impose open boundary conditions
and set the electric field at the left boundary to a
constant external field $\varepsilon$, with
$L_{-1}=\varepsilon$ \cite{ref28,ref57,ref12,ref60}.
Iterating Gauss's law from the left boundary then gives}

\begin{equation}
L_{n} = \varepsilon + \sum_{l = 0}^{n}
\left[ \phi_{l}^{\dagger}\phi_{l}
- \frac{1 - ( - 1)^{l}}{2} \right].
\end{equation}

{Here, $l$ is the summation index over the lattice sites
from the left boundary to site $n$.}

Therefore, the external field $\varepsilon$ is not introduced as an
independent potential-energy term. Instead, it enters the Hamiltonian
through the link and the corresponding electric-field
energy \cite{ref28,ref57,ref12}. It modifies the background electric field
configuration and thereby affects both the ground-state structure and
the real-time dynamics.

A further gauge transformation can absorb the link variable $U_{n}$ in
the hopping term \cite{ref29,ref30,ref57,ref12}. The Hamiltonian can then be written
in a purely fermionic form as

\begin{eqnarray}
H &=& -\frac{i}{2a}\sum_{n = 0}^{N - 2}
\left( \phi_{n}^{\dagger}\phi_{n + 1} - H.c. \right)
+ m\sum_{n = 0}^{N - 1}( - 1)^{n}
\phi_{n}^{\dagger}\phi_{n} \nonumber\\
&&+ \frac{g^{2}a}{2}\sum_{n = 0}^{N - 2}
\left[ \varepsilon + \sum_{l = 0}^{n}
\left( \phi_{l}^{\dagger}\phi_{l}
- \frac{1 - ( - 1)^{l}}{2} \right) \right]^{2}.
\end{eqnarray}

This expression shows that, after the gauge field is eliminated, the electric field energy becomes a long-range interaction induced by the
charge distribution. The external field $\varepsilon$, together with the lattice charge distribution, determines the electric field on each
link and is a key ingredient of the strong-field dynamics studied in
this work.

\subsection{Qubit Encoding}

To implement the above model in digital quantum simulation, the
fermionic operators must be mapped to Pauli operators. In this work, we
use the Jordan-Wigner transformation \cite{ref37,ref38,ref39},

\begin{equation}
\phi_{n} =
\left[ \prod_{l = 0}^{n - 1}
\left( -i\sigma_{l}^{z} \right) \right]
\frac{\sigma_{n}^{x} - i\sigma_{n}^{y}}{2}.
\end{equation}

After the Jordan-Wigner mapping, the lattice Schwinger model can be
written as a Pauli Hamiltonian,

\begin{equation}
H = H_{\mathrm{kin}} + H_{m} + H_{E},
\end{equation}

where the kinetic term is

\begin{equation}
H_{\mathrm{kin}} = \frac{1}{4a}\sum_{n = 0}^{N - 2}
\left( \sigma_{n}^{x}\sigma_{n + 1}^{x}
+ \sigma_{n}^{y}\sigma_{n + 1}^{y} \right),
\end{equation}

the mass term is

\begin{equation}
H_{m} = \frac{m}{2}\sum_{n = 0}^{N - 1}
( - 1)^{n}\sigma_{n}^{z},
\end{equation}

and the electric-field energy term is

\begin{equation}
H_{E} = \frac{g^{2}a}{2}\sum_{n = 0}^{N - 2}
\left[ \varepsilon + \frac{1}{2}\sum_{l = 0}^{n}
\left( \sigma_{l}^{z} + ( - 1)^{l} \right) \right]^{2}.
\end{equation}

Since $H_{E}$ contains the cumulative charge from the left boundary to
the $n$-th lattice site, it generates long-range interactions in the
Pauli representation. This nonlocal structure is a direct consequence of
eliminating the gauge degrees of freedom in one spatial dimension, and it distinguishes the present model from ordinary nearest-neighbor spin
chains \cite{ref29,ref30,ref57,ref12}.

\subsection{Observables}

To characterize the nonequilibrium dynamics under a strong external
electric field, we mainly focus on the following observables.

The total charge operator is defined as

\begin{equation}
Q_{N} = \frac{1}{2}\sum_{n = 0}^{N - 1}
\left( \sigma_{n}^{z} + ( - 1)^{n} \right).
\end{equation}

The total charge is a conserved quantity associated with the $U(1)$
gauge symmetry \cite{ref53,ref11,ref12}. During real-time evolution, the conservation of the total charge can be used as a basic check of the
correctness of the initial state, the Gauss-law constraint, and the
time-evolution algorithm.

The local charge density on the $n$-th lattice site is defined as

\begin{equation}
q_{n}(t) = \frac{1}{2}\left( \langle\sigma_{n}^{z}(t)\rangle + ( - 1)^{n} \right).
\end{equation}

In order to display the physical charge distribution
more clearly, we also measure the local charge density on each spatial point, which is the summation of charge densities on neighboring staggered
lattice sites. For each
spatial point $i$, the spatial-point charge is then defined as

\begin{equation}
Q_i(t) = q_{2i-2}(t)+q_{2i-1}(t).
\end{equation}

This observable is used to characterize the spatial separation of positive and negative charges induced by the external electric field.

The corresponding total electric-field energy is evaluated as the
expectation value of the electric-field-energy operator in Eq.~(12),
\begin{equation}
H_E(t) =
\frac{g^2 a}{2}
\sum_{n=0}^{N-2}
\left\langle
\left[
\varepsilon
+
\frac{1}{2}
\sum_{l=0}^{n}
\left(
\sigma_l^z + (-1)^l
\right)
\right]^2
\right\rangle_t .
\end{equation}
Here, $\langle\cdots\rangle_t$ denotes the expectation value with
respect to the time-evolved state $|\psi(t)\rangle$.

To quantify the stability of the initial vacuum state under the external
field, we define the vacuum-state fidelity as \cite{ref3,ref9,ref49,ref12}

\begin{equation}
P_{\mathrm{vac}}(t) = |\langle\psi_{0}|\psi(t)\rangle|^{2},
\end{equation}

where $\psi_{0}$ is the zero-field ground state and $\psi(t)$ is the state evolved after the electric-field quench.

For the static scan over the external field, we also introduce the
chiral condensate density \cite{ref27,ref28,ref31,ref32,ref33,ref34,ref35,ref9,ref36,ref10,ref59,ref60},

\begin{equation}
\Gamma = \langle\bar{\psi}\psi\rangle =
\frac{1}{2Na}\sum_{n = 0}^{N - 1}
( - 1)^{n}\langle\sigma_{n}^{z}\rangle,
\end{equation}

which is used to identify field-induced ground-state reconstruction in
finite-size systems.

\section{Numerical Methods and Quantum-Simulation Protocol}

\subsection{Exact Diagonalization and Benchmark
Calculations}

For a given system size $N$, lattice spacing $a$, fermion mass
$m$, gauge coupling $g$, and external-field strength
$\varepsilon$, we construct the sparse-matrix Hamiltonian of the
lattice Schwinger model and use exact diagonalization (ED) to obtain the
ground state, the low-energy spectrum, and benchmark real-time dynamics.
These results are used to assess the accuracy of the quantum algorithms.

\subsection{VQE Ground-State Preparation and VQD Excited-State
Estimation}

The initial state for the real-time evolution is taken as the ground
state of the zero-field Hamiltonian $H_{\varepsilon = 0}$. In this
work, we use the variational quantum eigensolver (VQE) to prepare this
initial vacuum state \cite{ref73,ref74,ref75,ref76}. For a trial state $\psi(\theta)$ generated by a parameterized
quantum circuit, {where
$\theta=(\theta_1,\theta_2,\ldots)$ denotes the set of
variational circuit parameters,} VQE approximates the
ground-state energy and wave function by minimizing the
energy expectation value

\begin{equation}
E(\theta) = \langle\psi(\theta) | H | \psi(\theta)\rangle.
\end{equation}

We employ the hardware-efficient RealAmplitudes
parameterized circuit and use the
{Sequential Least Squares Programming (SLSQP) algorithm}
as the classical optimizer \cite{ref75,ref82,ref83}. To reduce the influence of local minima, multiple random initial-parameter restarts
are used during the optimization \cite{ref84}. The accuracy of the VQE-prepared initial state is evaluated by comparing the ground-state
energy and wave-function fidelity with the ED results. In addition to
the zero-field initial-state preparation, VQE is also used in the static
external-field scan to estimate the ground-state energy
$E_{0}(\varepsilon)$ and the corresponding chiral condensate at
different external-field strengths.

To further compare the low-energy structure, the first excited-state energy is estimated using the variational quantum deflation (VQD) method
with an orthogonality penalty \cite{ref85}. After obtaining the approximate
VQE ground state $\psi_{0}^{VQE}$, the variational objective function
for the first excited state is taken as

\begin{equation}
C_{1}(\theta) = \langle\psi(\theta) | H | \psi(\theta)\rangle + \beta\left| \langle\psi_{0}^{VQE} | \psi(\theta)\rangle \right|^{2},
\end{equation}

where $\beta$ is the orthogonality-penalty coefficient. The second term suppresses the overlap between the candidate state and the VQE
ground state, thereby driving the optimization toward the
first-excited-state subspace. After the optimization, the first
excited-state energy is evaluated as

\begin{equation}
E_{1}^{VQD} = \langle\psi_{1}^{VQD} | H | \psi_{1}^{VQD}\rangle,
\end{equation}

and its accuracy is assessed by comparison with the ED result
$E_{1}^{ED}$. In the external-field scan, the optimized parameters
from the previous field point are used as the initial parameters for the
next field point to improve optimization stability. Thus, the
quantum-algorithm results in the static external-field scan consist of
two parts: VQE provides the ground-state energy and the chiral
condensate, while VQD provides the first excited-state energy.

\subsection{Second-Order Trotter-Suzuki Real-Time
Evolution}

After obtaining the zero-field ground state, we study the real-time dynamics after the external electric field is suddenly switched on. The
initial state is taken as the ground state at $\varepsilon = 0$. After the quench, the system evolves under the Hamiltonian
$H_{\varepsilon}$,

\begin{equation}
\left| \psi(t)\rangle = e^{- iH_{\varepsilon}t} \right|\psi_{0}\rangle.
\end{equation}

Since the Hamiltonian is decomposed into several noncommuting terms, the full time-evolution operator cannot be implemented directly \cite{ref77,ref78,ref79,ref80,ref81}. If the total Hamiltonian is written as a sum of $L$ terms, $H = \sum_{j = 1}^{L}H_{j}$, and the total evolution time $T$ is divided into $r$ time steps, the second-order Trotter-Suzuki decomposition is given by

\begin{equation}
e^{- iHT} \approx \left[ \prod_{j = 1}^{L}e^{- iH_{j}\Delta t/2}\prod_{j = L}^{1}e^{- iH_{j}\Delta t/2} \right]^{r},
\end{equation}

where $\Delta t = T/r$. The local error of each time step is
$O\left( \Delta t^{3} \right)$, and the global error is second order in $\Delta t$ for a fixed total evolution time. Compared with the
first-order decomposition, the symmetric second-order formula provides
higher accuracy for the same time step, at the cost of an increased circuit
depth \cite{ref77,ref78,ref79,ref80,ref81}.

In the numerical implementation, the full Pauli Hamiltonian is passed to
a second-order Suzuki-Trotter synthesizer, and Suzuki-Trotter (order=2,
reps=1) is used to generate the evolution circuit for a single time
step. The main time step is chosen as $\Delta t = 0.1$, and the total
evolution time is taken in the range $T/a \in [ 0,12]$. {A time-step convergence check using $\Delta t=0.2$, $0.1$,
and $0.05$ confirms the expected second-order convergence and
shows that $\Delta t=0.1$ provides sufficient accuracy for the
dynamical observables considered here.}
The representative external-field strengths in the dynamical
calculations are selected according to the observable under
consideration. In particular, the spatial-point charge distribution and
electric-field energy response are mainly studied for
$\varepsilon = 0.5,1.0,1.5,2.0$, while the analysis of the
vacuum-state fidelity and the effective decay rate is further extended
to stronger external fields. The numerical implementation is based on
Qiskit and the standard scientific Python packages \cite{ref82,ref83,ref86,ref87,ref88}. The current simulation is performed on classical computers without considering quantum errors. 
In the present encoding, the circuit width is equal to the lattice
size $N$. For the
two-layer RealAmplitudes ansatz used in most dynamical calculations,
the circuit depths are 24 and 36 for $N=8$ and $N=12$, respectively.
After decomposition into elementary one- and two-qubit gates, a single
second-order Suzuki--Trotter step has depths of approximately 287 and
455, respectively, and the total real-time evolution depth increases
linearly with the number of Trotter steps.

\section{Numerical Results}

\begin{table*}[!htb]
    \centering
    \caption{Comparison between VQE and ED for ground state preparation}
    \begin{tabular}{c|cccc}
        \hline
        System parameters & ED energy & VQE energy & Relative error & Fidelity \\
        \hline
        $N = 8$, $a = m = g = 1$, $\varepsilon = 0$ & $-4.63805774$ & $-4.63766032$ & $8.569\times 10^{-5}$ & $99.9931\%$ \\
        \hline
    \end{tabular}
    \label{Tab:state-compare}
\end{table*}

\begin{figure*}[!htb]
    \centering
    \includegraphics[width=0.8\textwidth]{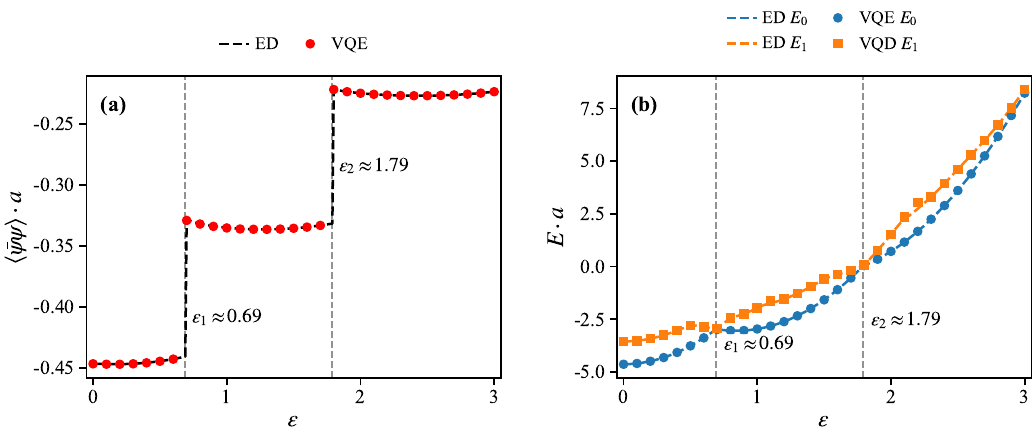}
    \caption{Static observables as functions of the external-field strength $\varepsilon$. The left figure shows the chiral condensate of the ground state; the right figure shows the energy of the ground state and the first excited state.}
    \label{Fig:static-obs}
\end{figure*}

\begin{figure}[!htb]
    \centering    
    \includegraphics[width=0.4\textwidth]{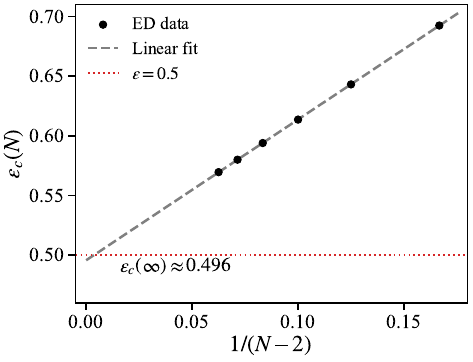}
    \caption{{Finite-size critical field $\varepsilon_c(N)$ as a
function of the boundary-shifted inverse size $1/(N-2)$. }}
    \label{fig:epsilonvsN}
\end{figure}

The main results are summarized as follows. For the system with
$N = 8$ and $a = m = g = 1$, VQE prepares the ground state with high accuracy. We scan over the external field strength and calculate the chiral condensate and energy gap. These features are consistent with the flip of the finite-volume ground state. We then investigate the electric-field quench of the zero-field ground state. The second-order Trotter-Suzuki evolution preserves the total charge and total energy and shows good agreement with the ED results. The external electric field drives oscillations of the spatial point charges near the
open boundaries; the vacuum-state fidelity exhibits an effective
exponential decay within the early-time window, and the electric field energy shows quasiperiodic behavior, reflecting finite-size energy
redistribution between the electric-field energy term and the fermionic
sector. 

\subsection{Ground-State Preparation and Study of the Critical Field Strength}

We first examine the VQE preparation of the zero-field ground state for
$N = 8$, $a = m = g = 1$, and $\varepsilon = 0$. The results are
shown in Table \ref{Tab:state-compare}. The fidelity is defined as $F=|\langle\psi_{0}^{\mathrm{ED}}|\psi_{0}^{\mathrm{VQE}}\rangle|^{2}$. These results indicate that the chosen parameterized
circuit can approximate the zero-field ground state with high accuracy
at the present system size, providing a reliable initial state for the
subsequent electric-field-quench dynamics.

We then turn on the external field, prepare the ground and first excited states.
We calculate the chiral condensate $\langle\bar\psi\psi\rangle$ of the
ground state, {as shown in Figure \ref{Fig:static-obs}(a), and the
ground state energy $E_0$ and the first excited state energy $E_1$, as
shown in Figure \ref{Fig:static-obs}(b).}

The chiral condensate closely resembles a step function, and the jumps occur when the energy gap between the ground state and the first excited state vanishes. This result indicates that the external electric field can modify the vacuum structure of the finite system, driving the ground state from one to another. This phenomenon corresponds to a first-order phase transition at zero temperature, as observed in various studies of 1+1D models \cite{ref28,ref32,ref59}. The VQE results are in overall agreement with the ED results.

\begin{figure*}[!htb]
    \centering
    \includegraphics[width=0.8\textwidth]{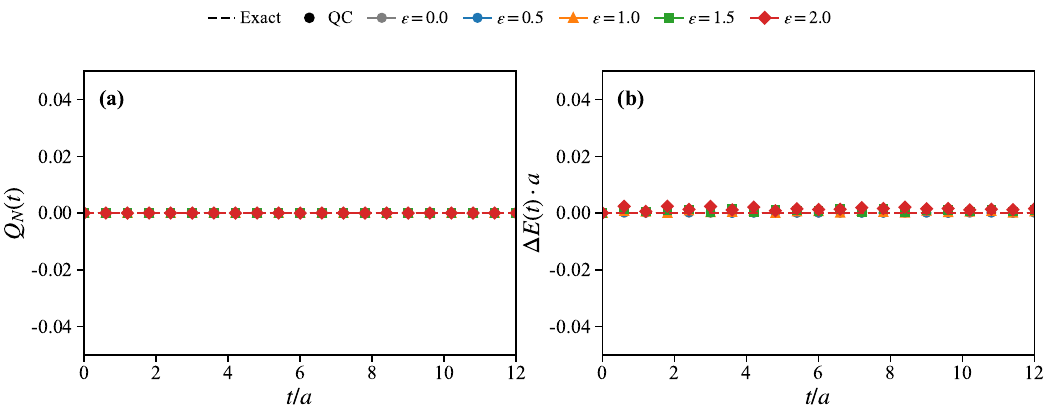}
    \caption{Time evolution of (left) the total charge $Q_N(t)$ and (right)
the total energy shift $\Delta E(t) = E(t) - E(0)$. In both panels, dashed lines denote ED results, while dots denote results obtained from the quantum algorithm.}
    \label{fig:qcheck}
\end{figure*}

It should be emphasized that the external-field scan discussed here
is performed in finite-size systems, so the location of the low-energy
spectral reconstruction depends on the system size. For each lattice
size $N$, we define the energy gap as
$\Delta(\varepsilon,N)=E_1(\varepsilon,N)-E_0(\varepsilon,N)$,
and denote by $\varepsilon_c(N)$ the lowest positive external-field strength $\varepsilon$ at which $\Delta(\varepsilon,N) = 0$. For a simple estimation, we assume the following picture:
at $\varepsilon < \varepsilon_c$, the ground state has a trivial
particle distribution while the first excited state has an antiparticle
at the left boundary, screening the external field. In this configuration,
the antiparticle corresponds to a local charge $q_1=-1$ on the odd
staggered site $n=1$; according to Gauss's law, $L_0$ remains equal to
$\varepsilon$, whereas the remaining $N-2$ internal links,
$L_1,\ldots,L_{N-2}$, are shifted from $\varepsilon$ to
$\varepsilon-1$. Then the energy gap comes mostly from the mass of the
particles and the difference in the electric field energy, $\Delta(\varepsilon,N)\approx \frac{1}{2}(N-2)(\varepsilon-1)^2 + m-\frac{1}{2}(N-2)\varepsilon^2=\frac{1}{2}(N-2)(1-2\varepsilon)+m$. So the critical field $\varepsilon_c(N)\approx \frac{m}{(N-2)}+\frac{1}{2}$. We plot $\varepsilon_c(N)$ as a function of $1/(N-2)$
in Figure \ref{fig:epsilonvsN}. The data exhibit an approximately linear dependence on
this variable, and the linear extrapolation gives
$\varepsilon_c(\infty)\approx0.496$. The infinite-size value $\frac{1}{2}$ can also be derived from Coleman's analysis of the massive Schwinger model: $\theta=\pi$ is the special point associated with the first-order
transition \cite{ref28}. In the equivalent background-field description,
$\varepsilon=\theta/(2\pi)$. Therefore, the Coleman
point $\theta=\pi$ corresponds to the theoretical reference value
$\varepsilon=1/2$.

\subsection{Dynamical Evolution under an External Field}
We proceed to the main part of this work. We prepare the ground state in the absence of an external field, then turn the field on at $t=0$ and study the time evolution of the state.

First, we test the conservation laws after the electric-field quench. For different external-field strengths, the total charge
$Q_N(t)$ {remains extremely close to zero} throughout
the time evolution. This indicates that the second-order Trotter-Suzuki evolution preserves the $U(1)$ charge as is expected.

We also calculate the total energy shift, $\Delta E(t) = E(t) - E(0)$.
$\Delta E(t)$ remains very close to zero throughout the simulated time interval. This shows that, with the chosen time step, the time evolution is numerically stable. The results are shown in Figure \ref{fig:qcheck}.

\begin{figure*}[!htb]
    \centering
    \includegraphics[width=0.8\textwidth]{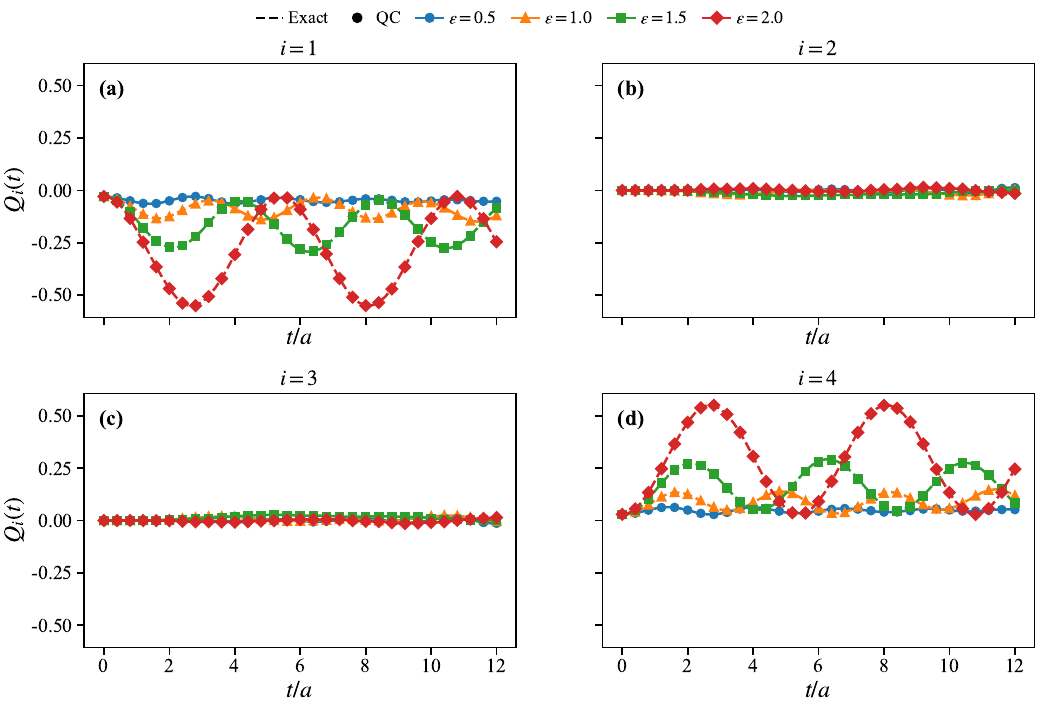}
    \caption{Time evolution of the spatial charge distribution}
    \label{fig:qspatial}
\end{figure*}

\begin{figure*}[!htb]
    \centering
    \includegraphics[width=0.8\textwidth]{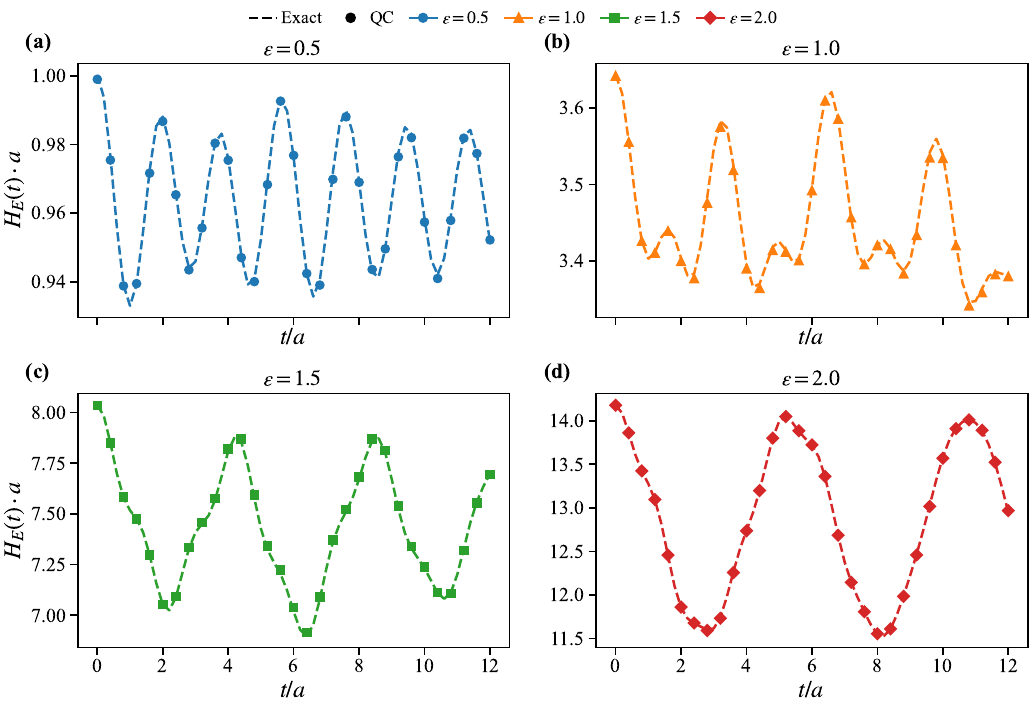}
    \caption{Time evolution of the total electric field
energy $H_{E}(t)$ under different external-field strengths}
    \label{fig:He}
\end{figure*}

Although the total charge is conserved, the spatial charge distribution $Q_i(t)$ undergoes a strong rearrangement, as displayed in Figure \ref{fig:qspatial}. After the field quench, a charge-anticharge pair is periodically produced and annihilated on the boundary of the system. The period and amplitude of the oscillation increase with the external field strength. {The response is strongly enhanced near the open boundaries,
whereas $Q_2(t)$ and $Q_3(t)$ exhibit only weaker oscillations around
zero. This behavior can be understood as the screening effect: a pair of charges on the boundary provides a more
effective screening of the external background field.} As a related quantity, we also calculate the time evolution of total electric field energy {$H_E(t)$ for different $\varepsilon$, as shown in}  Figure \ref{fig:He}. When the charge pair is produced, energy is transferred from the electric field to the fermions, and when it is annihilated, the energy returns to the field. However, $H_E(t)$ does not show as simple a periodic pattern as $Q_i(t)$. 
In the staggered-fermion description, the charge at one
physical spatial point is obtained by combining two neighboring
staggered-site contributions as
$Q_i(t)=q_{2i-2}(t)+q_{2i-1}(t)$, with the corresponding site-resolved
results shown in the Appendix. The electric-field energy $H_E(t)$,
on the other hand, is determined by the sum of squares of electric fields over the whole
chain and therefore need not follow the same simple oscillation pattern as $Q_i(t)$.

It is also worth noting that the dependence of the period and amplitude
on $\varepsilon$ appears to be smooth,
{unlike the step-like behavior of the static ground-state chiral
condensate shown in Figure \ref{Fig:static-obs}(a).}
This is not surprising, because we can see from Figure
\ref{Fig:static-obs}{(b)} that the eigen energies' dependence on
$\varepsilon$ is actually smooth. 

To characterize the system's rate of deviation from the initial state, we calculate the vacuum-state fidelity
$P_{\mathrm{vac}}(t) = \left| \langle\psi_{0} \right|\psi(t)\rangle|^{2}$. This quantity describes the probability that the system remains in the
initial vacuum state under the driving of the strong external electric field, and is used to characterize the field-induced deviation from the
vacuum state.

\begin{figure*}[!htb]
    \centering
    \includegraphics[width=0.8\textwidth]{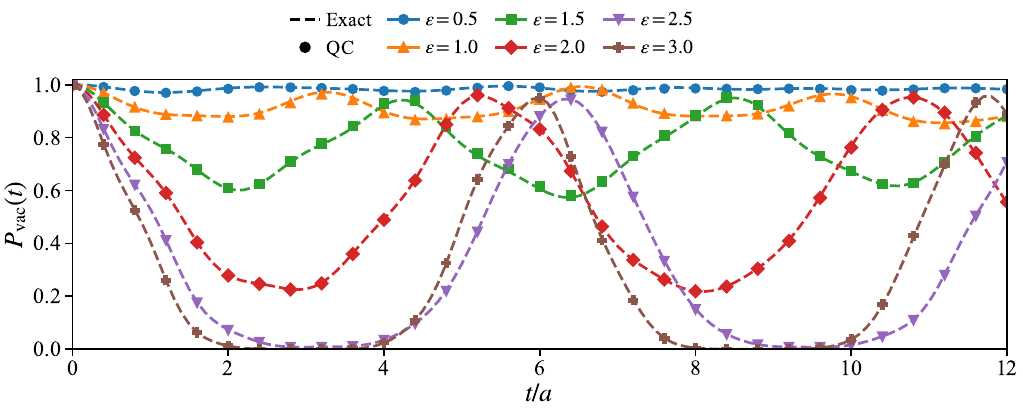}
    \caption{Time evolution of the vacuum-state fidelity under different external field strengths}
    \label{fig:fidelity}
\end{figure*}

Figure \ref{fig:fidelity} shows that, for weak external fields, the vacuum-state fidelity remains close to unity. As the external field is increased, the
system deviates from the zero-field vacuum more rapidly, indicating that a stronger external field enhances the nonequilibrium response of the
system. Since the present study concerns a finite closed system,
$P_{\mathrm{vac}}(t)$ does not exhibit strictly monotonic decay, but instead
shows finite-size revivals and quasiperiodic oscillations similar to $H_E$. Again, the quantum algorithm results agree well with the ED results, demonstrating that the evolution protocol captures the vacuum response of the finite-lattice system.

To further quantify the initial decrease of the vacuum-state fidelity under different external fields, we perform an effective exponential fit to $P_{\mathrm{vac}}(t;\varepsilon)$ in the early-time window $0 \leq t/a \leq 1.0$:
$P_{\mathrm{vac}}(t;\varepsilon) \simeq A(\varepsilon)\exp\left[ - \gamma_{\mathrm{eff}}(\varepsilon)t \right].$
Equivalently, we fit $-\ln P_{\mathrm{vac}}(t;\varepsilon) = \gamma_{\mathrm{eff}}(\varepsilon)t + c(\varepsilon)$, and define the slope as the effective decay rate $\gamma_{\mathrm{eff}}$. Within this early-time window, the numerical results show that $-\ln P_{\mathrm{vac}}(t;\varepsilon)$ is approximately linear in time, indicating that the initial decrease of the vacuum-state fidelity can be characterized by an exponential form.

\begin{figure}[!htb]
    \centering
    \includegraphics[width=0.4\textwidth]{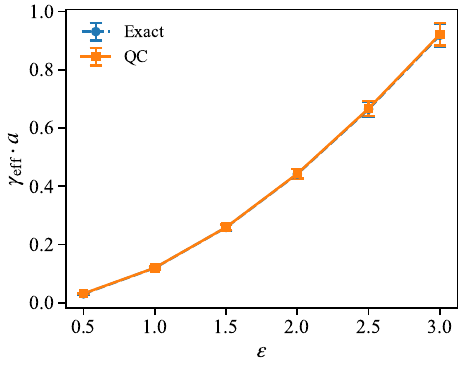}
    \caption{Effective decay rate $\gamma_{\mathrm{eff}}$ obtained by
fitting}
    \label{fig:decayrate}
\end{figure}

Figure \ref{fig:decayrate} shows the dependence of $\gamma_{\mathrm{eff}}$ on the
external-field strength $\varepsilon$ extracted from the above fitting
procedure. The effective decay rate increases monotonically with
$\varepsilon$, indicating that a stronger external field drives the system away from the initial vacuum state more rapidly.

As a system-size check, Appendix A also presents dynamical comparison
results for the $N = 12$ lattice system. Similar to the $N = 8$ case, the spatial charge distribution in the $N = 12$ system
still shows boundary charge separation induced by the external electric
field, indicating that this phenomenon is not a numerical artifact
specific to the small $N = 8$ system. Under the same external field
strength, $P_{\mathrm{vac}}(t)$ in the $N = 12$ system is more sensitive to
the external field and deviates from its initial value more rapidly.
This may be related to the denser low-energy spectrum and the larger
number of excited states participating in the quench dynamics in the
larger system.

\section{Summary and Outlook}

In this work, we studied the static ground-state structure and nonequilibrium dynamics of the finite-lattice Schwinger model under a constant external electric field. We used VQE and VQD to calculate the low-energy states and employed a second-order Trotter-Suzuki decomposition to simulate the real-time evolution after an electric-field quench. The quantum-algorithm results were compared with exact diagonalization throughout the calculation.

The external-field scan shows step-like changes in the chiral condensate near the field strengths where the ground state and the first excited state become closest in energy. The corresponding characteristic field strength exhibits a clear finite-size dependence. After the electric-field quench, the total charge and total energy remain approximately conserved. The system exhibits boundary charge separation, quasiperiodic electric-field-energy redistribution, and a field-dependent decrease of the vacuum-state fidelity. The site-resolved charge dynamics and the $N=12$ results further show that the main dynamical features persist for a larger lattice.

These results demonstrate that VQE-based state preparation combined with digital real-time evolution can reproduce the main static and dynamical responses of the finite-lattice Schwinger model. Future work may extend the present study to larger systems, improved variational ansatzes, noisy quantum hardware, and time-dependent external fields.

\section*{Acknowledgments}
This work is supported by the National Natural Science Foundation of China under Grant Nos. 12547109, 12525508, and 12475139.

\appendix

\section{Site-Resolved Charge Dynamics and the $N=12$ Lattice}

The real-time dynamics in the main text mainly focus on the $N=8$ lattice and use the spatial-point charge $Q_i(t)$ to show the field-induced charge separation. In this appendix, we provide the corresponding site-resolved charge dynamics and the $N=12$ lattice check. The line and marker conventions are the same as in the main text.

\begin{figure*}[!htb]
\centering
\includegraphics[width=0.8\textwidth]{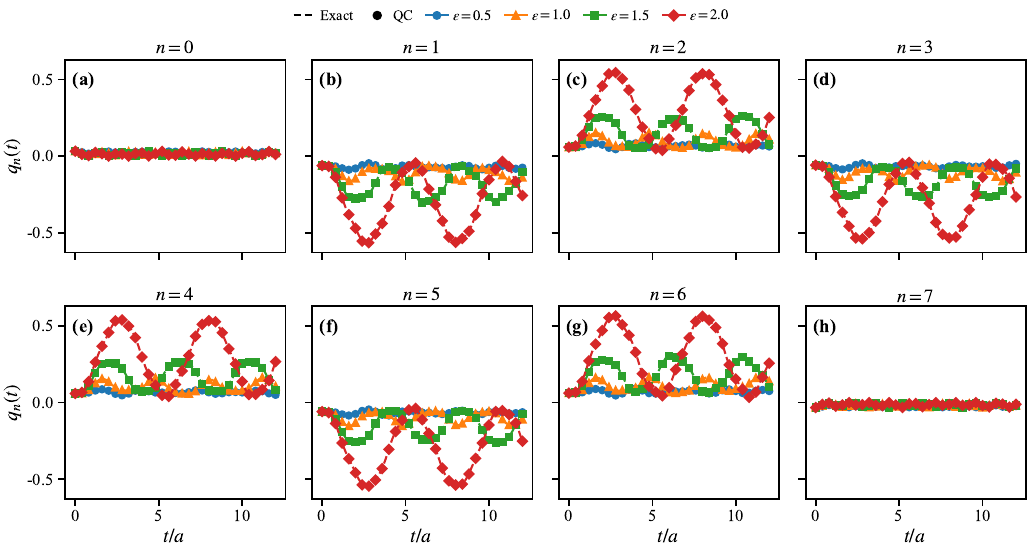}
\caption{Time evolution of the site-resolved charge $q_n(t)$ in the $N=8$ lattice.}
\label{fig:site_charge_N8}
\end{figure*}

Figure~\ref{fig:site_charge_N8} shows the time evolution of the staggered-site charge $q_n(t)$ for the $N=8$ lattice. Compared with the spatial-point charge $Q_i(t)$ in the main text, the site-resolved charge contains more detailed staggered-component oscillations. The dominant response still appears near the open boundaries.

\begin{figure*}[!htb]
\centering
\includegraphics[width=0.8\textwidth]{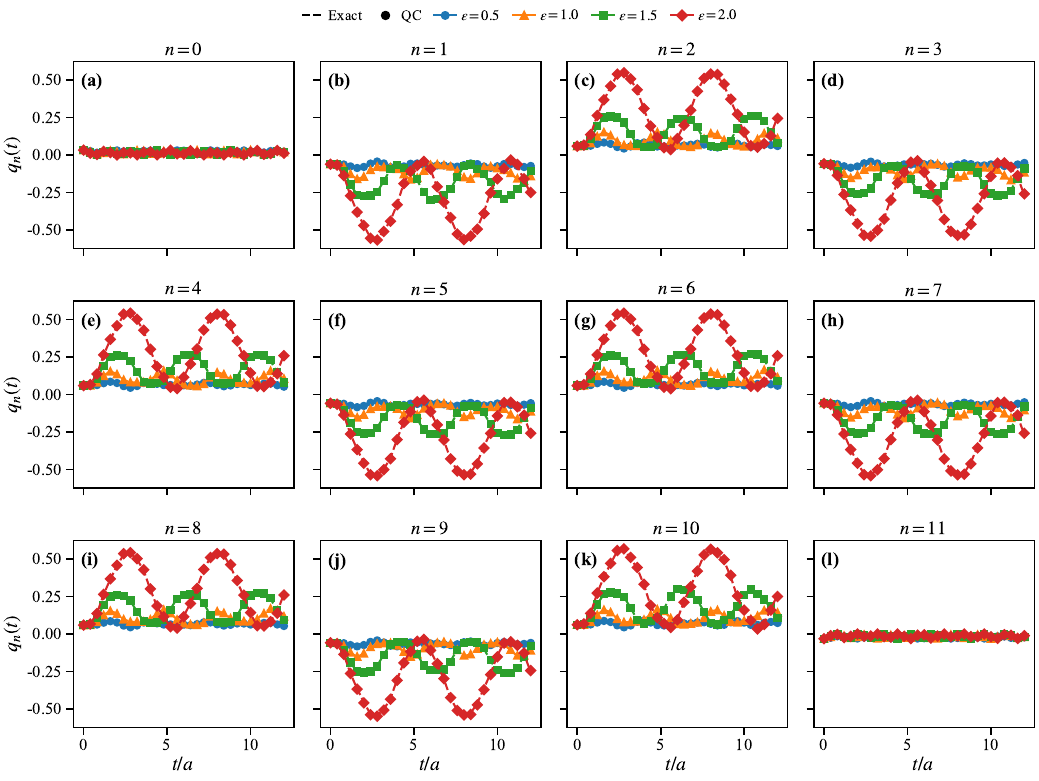}
\caption{Time evolution of the site-resolved charge $q_n(t)$ in the $N=12$ lattice.}
\label{fig:site_charge_N12}
\end{figure*}

Figure~\ref{fig:site_charge_N12} shows the corresponding site-resolved charge dynamics for the $N=12$ lattice. The larger system contains more local oscillatory components, but the boundary-dominated response remains similar to the $N=8$ case.

\begin{figure*}[!htb]
\centering
\includegraphics[width=0.8\textwidth]{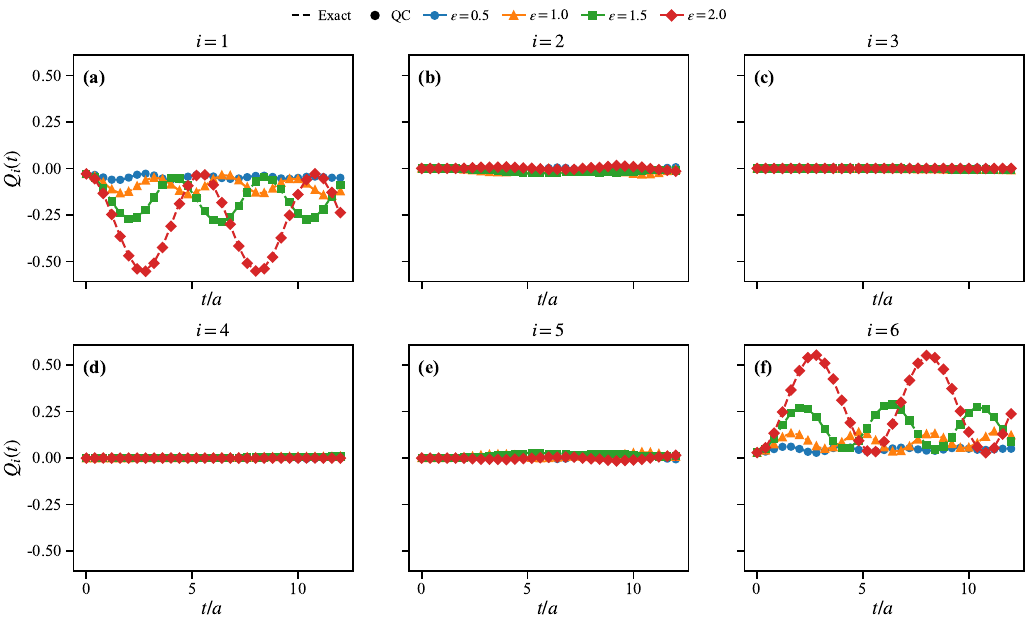}
\caption{Time evolution of the spatial-point charge $Q_i(t)$ in the $N=12$ lattice.}
\label{fig:spatial_charge_N12}
\end{figure*}

For comparison with the spatial-point observable used in the main text, Figure~\ref{fig:spatial_charge_N12} shows $Q_i(t)$ for the $N=12$ lattice. The result shows the same qualitative boundary charge separation as in the $N=8$ lattice.

\begin{figure*}[!htb]
\centering
\includegraphics[width=0.8\textwidth]{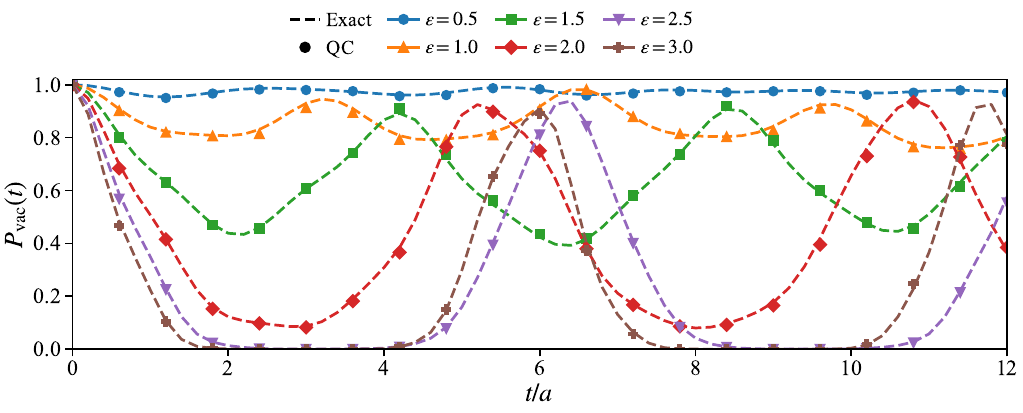}
\caption{Time evolution of the vacuum-state fidelity $P_{\mathrm{vac}}(t)$ in the $N=12$ lattice.}
\label{fig:fidelity_N12}
\end{figure*}

We also compare the vacuum-state fidelity for the $N=12$ lattice. As shown in Figure~\ref{fig:fidelity_N12}, $P_{\mathrm{vac}}(t)$ deviates from its initial value more rapidly than in the $N=8$ case under the same external-field strength. This behavior may be related to the denser low-energy spectrum and the larger number of excited states involved in the quench dynamics.

Overall, the $N=8$ and $N=12$ results show that the boundary charge separation and the deviation from the initial vacuum state persist as the system size is increased.


\clearpage

\section*{References}
\begin{thebibliography}{99}
\small

\bibitem{ref1}
F. Sauter, Z. Phys., \textbf{69}: 742 (1931)

\bibitem{ref2}
W. Heisenberg and H. Euler, Z. Phys., \textbf{98}: 714 (1936)

\bibitem{ref3}
J. Schwinger, Phys. Rev., \textbf{82}: 664 (1951)

\bibitem{ref4}
E. Brezin and C. Itzykson, Phys. Rev. D, \textbf{2}: 1191 (1970)

\bibitem{ref5}
V. S. Popov, Sov. Phys. JETP, \textbf{34}: 709 (1972)

\bibitem{ref6}
A. Ringwald, Phys. Lett. B, \textbf{510}: 107 (2001)

\bibitem{ref7}
R. Alkofer, M. B. Hecht, C. D. Roberts, \textit{et al.}, Phys. Rev. Lett., \textbf{87}: 193902 (2001)

\bibitem{ref8}
G. V. Dunne, Eur. Phys. J. D, \textbf{55}: 327 (2009)

\bibitem{ref9}
B. Buyens, J. Haegeman, F. Hebenstreit, \textit{et al.}, Phys. Rev. D, \textbf{96}: 114501 (2017)

\bibitem{ref10}
T. V. Zache, N. Mueller, J. T. Schneider, \textit{et al.}, Phys. Rev. Lett., \textbf{122}: 050403 (2019)

\bibitem{ref11}
A. F. Shaw, P. Lougovski, J. R. Stryker, \textit{et al.}, Quantum, \textbf{4}: 306 (2020)

\bibitem{ref12}
L. Nagano, A. Bapat, and C. W. Bauer, Phys. Rev. D, \textbf{108}: 034501 (2023)

\bibitem{ref13}
K. G. Wilson, Phys. Rev. D, \textbf{10}: 2445 (1974)

\bibitem{ref14}
J. Kogut and L. Susskind, Phys. Rev. D, \textbf{11}: 395 (1975)

\bibitem{ref15}
J. B. Kogut, Rev. Mod. Phys., \textbf{51}: 659 (1979)

\bibitem{ref16}
M. Creutz, \textit{Quarks, Gluons and Lattices} (Cambridge: Cambridge University Press, 1983)

\bibitem{ref17}
I. Montvay and G. M\"unster, \textit{Quantum Fields on a Lattice} (Cambridge: Cambridge University Press, 1994)

\bibitem{ref18}
M. Troyer and U.-J. Wiese, Phys. Rev. Lett., \textbf{94}: 170201 (2005)

\bibitem{ref19}
P. de Forcrand, Simulating QCD at finite density, PoS, \textbf{LAT2009}: 010 (2009)

\bibitem{ref20}
G. Aarts, Complex Langevin dynamics and other approaches at finite chemical potential, PoS, \textbf{LATTICE2012}: 017 (2012)

\bibitem{ref21}
R. P. Feynman, Int. J. Theor. Phys., \textbf{21}: 467 (1982)

\bibitem{ref22}
S. Lloyd, Science, \textbf{273}: 1073 (1996)

\bibitem{ref23}
S. P. Jordan, K. S. M. Lee, and J. Preskill, Science, \textbf{336}: 1130 (2012)

\bibitem{ref24}
J. Preskill, Quantum, \textbf{2}: 79 (2018)

\bibitem{ref25}
J. Schwinger, Phys. Rev., \textbf{128}: 2425 (1962)

\bibitem{ref26}
J. H. Lowenstein and J. A. Swieca, Ann. Phys. (N.Y.), \textbf{68}: 172 (1971)

\bibitem{ref27}
S. Coleman, R. Jackiw, and L. Susskind, Ann. Phys. (N.Y.), \textbf{93}: 267 (1975)

\bibitem{ref28}
S. Coleman, Ann. Phys. (N.Y.), \textbf{101}: 239 (1976)

\bibitem{ref29}
A. Casher, J. Kogut, and L. Susskind, Phys. Rev. D, \textbf{10}: 732 (1974)

\bibitem{ref30}
T. Banks, L. Susskind, and J. Kogut, Phys. Rev. D, \textbf{13}: 1043 (1976)

\bibitem{ref31}
C. J. Hamer, Z. Weihong, and J. Oitmaa, Phys. Rev. D, \textbf{56}: 55 (1997)

\bibitem{ref32}
T. Byrnes, P. Sriganesh, R. J. Bursill, \textit{et al.}, Phys. Rev. D, \textbf{66}: 013002 (2002)

\bibitem{ref33}
M. C. Ba\~nuls, K. Cichy, J. I. Cirac, \textit{et al.}, JHEP, \textbf{11}: 158 (2013)

\bibitem{ref34}
B. Buyens, J. Haegeman, K. Van Acoleyen, \textit{et al.}, Phys. Rev. Lett., \textbf{113}: 091601 (2014)

\bibitem{ref35}
B. Buyens, F. Verstraete, and K. Van Acoleyen, Phys. Rev. D, \textbf{94}: 085018 (2016)

\bibitem{ref36}
L. Funcke, K. Jansen, and S. K\"uhn, Phys. Rev. D, \textbf{101}: 054507 (2020)

\bibitem{ref37}
P. Jordan and E. Wigner, Z. Phys., \textbf{47}: 631 (1928)

\bibitem{ref38}
M. A. Nielsen and I. L. Chuang, \textit{Quantum Computation and Quantum Information} (Cambridge: Cambridge University Press, 2000)

\bibitem{ref39}
G. Ortiz, J. E. Gubernatis, E. Knill, \textit{et al.}, Phys. Rev. A, \textbf{64}: 022319 (2001)

\bibitem{ref40}
I. M. Georgescu, S. Ashhab, and F. Nori, Rev. Mod. Phys., \textbf{86}: 153 (2014)

\bibitem{ref41}
U.-J. Wiese, Ann. Phys. (Berlin), \textbf{525}: 777 (2013)

\bibitem{ref42}
E. Zohar, J. I. Cirac, and B. Reznik, Rep. Prog. Phys., \textbf{79}: 014401 (2016)

\bibitem{ref43}
M. C. Ba\~nuls, R. Blatt, J. Catani, \textit{et al.}, Eur. Phys. J. D, \textbf{74}: 165 (2020)

\bibitem{ref44}
C. W. Bauer, Z. Davoudi, A. B. Balantekin, \textit{et al.}, PRX Quantum, \textbf{4}: 027001 (2023)

\bibitem{ref45}
D. Banerjee, M. Dalmonte, M. M\"uller, \textit{et al.}, Phys. Rev. Lett., \textbf{109}: 175302 (2012)

\bibitem{ref46}
E. Zohar, J. I. Cirac, and B. Reznik, Phys. Rev. Lett., \textbf{109}: 125302 (2012)

\bibitem{ref47}
D. Banerjee, M. B\"ogli, M. Dalmonte, \textit{et al.}, Phys. Rev. Lett., \textbf{110}: 125303 (2013)

\bibitem{ref48}
E. Zohar, A. Farace, B. Reznik, \textit{et al.}, Phys. Rev. A, \textbf{95}: 023604 (2017)

\bibitem{ref49}
E. A. Martinez, C. A. Muschik, P. Schindler, \textit{et al.}, Nature, \textbf{534}: 516 (2016)

\bibitem{ref50}
N. Klco, E. F. Dumitrescu, A. J. McCaskey, \textit{et al.}, Phys. Rev. A, \textbf{98}: 032331 (2018)

\bibitem{ref51}
C. Kokail, C. Maier, R. van Bijnen, \textit{et al.}, Nature, \textbf{569}: 355 (2019)

\bibitem{ref52}
F. M. Surace, P. P. Mazza, G. Giudici, \textit{et al.}, Phys. Rev. X, \textbf{10}: 021041 (2020)

\bibitem{ref53}
J. R. Stryker, Phys. Rev. A, \textbf{99}: 042301 (2019)

\bibitem{ref54}
N. H. Nguyen, M. C. Tran, Y. Zhu, \textit{et al.}, PRX Quantum, \textbf{3}: 020324 (2022)

\bibitem{ref55}
W. A. de Jong, K. Lee, J. Mulligan, \textit{et al.}, Phys. Rev. D, \textbf{106}: 054508 (2022)

\bibitem{ref56}
X.-D. Xie, X. Guo, H. Xing, \textit{et al.}, Phys. Rev. D, \textbf{106}: 054509 (2022)

\bibitem{ref57}
M. Honda, E. Itou, Y. Kikuchi, \textit{et al.}, Phys. Rev. D, \textbf{105}: 014504 (2022)

\bibitem{ref58}
B. Chakraborty, M. Honda, T. Izubuchi, \textit{et al.}, Phys. Rev. D, \textbf{105}: 094503 (2022)

\bibitem{ref59}
T. Angelides, P. Naredi, A. Crippa, \textit{et al.}, npj Quantum Inf., \textbf{11}: 6 (2025)

\bibitem{ref60}
O. Kaikov, T. Saporiti, V. Sazonov, \textit{et al.}, Phys. Rev. A, \textbf{111}: 062202 (2025)

\bibitem{ref61}
G. Zhang, X. Guo, E. Wang, \textit{et al.}, Phys. Rev. D, \textbf{111}: 056031 (2025)

\bibitem{ref62}
B. Yang, H. Sun, R. Ott, \textit{et al.}, Nature, \textbf{587}: 392 (2020)

\bibitem{ref63}
Z.-Y. Zhou, G.-X. Su, J. C. Halimeh, \textit{et al.}, Science, \textbf{377}: 311 (2022)

\bibitem{ref64}
G.-X. Su, H. Sun, A. Hudomal, \textit{et al.}, Phys. Rev. Res., \textbf{5}: 023010 (2023)

\bibitem{ref65}
J. C. Halimeh, I. P. McCulloch, B. Yang, \textit{et al.}, PRX Quantum, \textbf{3}: 040316 (2022)

\bibitem{ref66}
W.-Y. Zhang, Y. Liu, Y. Cheng, \textit{et al.}, Nat. Phys., \textbf{21}: 155 (2025)

\bibitem{ref67}
G.-X. Su, J. J. Osborne, and J. C. Halimeh, PRX Quantum, \textbf{5}: 040310 (2024)

\bibitem{ref68}
J. Schuhmacher, G.-X. Su, J. J. Osborne, \textit{et al.}, arXiv:2505.20387

\bibitem{ref69}
R. Joshi, Y. Tian, K. Hemery, \textit{et al.}, arXiv:2604.07436

\bibitem{ref70}
A. Di Meglio, K. Jansen, I. Tavernelli, \textit{et al.}, PRX Quantum, \textbf{5}: 037001 (2024)

\bibitem{ref71}
J. C. Halimeh, N. Mueller, J. Knolle, \textit{et al.}, arXiv:2509.03586

\bibitem{ref72}
S. Gupta and L. Santos, Phys. Rev. A, \textbf{114}: 033321 (2026)

\bibitem{ref73}
A. Peruzzo, J. McClean, P. Shadbolt, \textit{et al.}, Nat. Commun., \textbf{5}: 4213 (2014)

\bibitem{ref74}
J. R. McClean, J. Romero, R. Babbush, \textit{et al.}, New J. Phys., \textbf{18}: 023023 (2016)

\bibitem{ref75}
A. Kandala, A. Mezzacapo, K. Temme, \textit{et al.}, Nature, \textbf{549}: 242 (2017)

\bibitem{ref76}
M. Cerezo, A. Arrasmith, R. Babbush, \textit{et al.}, Nat. Rev. Phys., \textbf{3}: 625 (2021)

\bibitem{ref77}
H. F. Trotter, Proc. Am. Math. Soc., \textbf{10}: 545 (1959)

\bibitem{ref78}
M. Suzuki, Commun. Math. Phys., \textbf{51}: 183 (1976)

\bibitem{ref79}
M. Suzuki, Phys. Lett. A, \textbf{146}: 319 (1990)

\bibitem{ref80}
D. W. Berry, A. M. Childs, R. Cleve, \textit{et al.}, Phys. Rev. Lett., \textbf{114}: 090502 (2015)

\bibitem{ref81}
A. M. Childs, Y. Su, M. C. Tran, \textit{et al.}, Phys. Rev. X, \textbf{11}: 011020 (2021)

\bibitem{ref82}
A. Javadi-Abhari, M. Treinish, K. Krsulich, \textit{et al.}, arXiv:2405.08810

\bibitem{ref83}
G. Aleksandrowicz, T. Alexander, P. Barkoutsos, \textit{et al.}, \textit{Qiskit: An Open-source Framework for Quantum Computing}, Zenodo (2019), doi:\href{https://doi.org/10.5281/zenodo.2562111}{10.5281/zenodo.2562111}

\bibitem{ref84}
J. R. McClean, S. Boixo, V. N. Smelyanskiy, \textit{et al.}, Nat. Commun., \textbf{9}: 4812 (2018)

\bibitem{ref85}
O. Higgott, D. Wang, and S. Brierley, Quantum, \textbf{3}: 156 (2019)

\bibitem{ref86}
P. Virtanen, R. Gommers, T. E. Oliphant, \textit{et al.}, Nat. Methods, \textbf{17}: 261 (2020)

\bibitem{ref87}
C. R. Harris, K. J. Millman, S. J. van der Walt, \textit{et al.}, Nature, \textbf{585}: 357 (2020)

\bibitem{ref88}
J. D. Hunter, Comput. Sci. Eng., \textbf{9}: 90 (2007)

\end{thebibliography}
\end{document}